\documentstyle[aps,epsfig,float,twocolumn,prl]{revtex}
\newcommand{\be}{\begin{equation}}
\newcommand{\ee}{\end{equation}}

\begin{document}
\twocolumn[\hsize\textwidth\columnwidth\hsize\csname @twocolumnfalse\endcsname
\draft
\title{High-Dimensional Diffusive Growth}
\author{M. B. Hastings$^{1,*}$ and Thomas C. Halsey$^2$}
\address{$^1$Physics Department, Jadwin Hall, Princeton University,
Princeton, NJ 08544\\
$^2$Corporate Strategic Research, ExxonMobil Research and Engineering,
Route 22 East, Annandale, New Jersey 08801
}
\date{14 August 2000}
\maketitle
\begin{abstract}
We consider a model of aggregation, both diffusion-limited and ballistic,
based on the Cayley tree.  Growth is from the
leaves of the tree towards the root, leading to non-trivial
screening and branch competition effects.  The model exhibits a phase transition
between ballistic and diffusion-controlled growth, with 
non-trivial corrections to cluster
size at the critical point.  Even in the ballistic regime,
cluster scaling is controlled by extremal statistics due to
the branching structure of the Cayley tree; it is the extremal
nature of the fluctuations that enables us to solve the model.
\end{abstract}
]
\section{Introduction}
Diffusion-limited aggregation (DLA)\cite{wittsand} is a model of central
importance in the field of fractal growth.  For the theorist, it
presents a model with very simple rules that gives rise to
complex, branching patterns; for the experimentalist, it
provides a model relevant for dielectric breakdown\cite{db},
viscous fingering\cite{vf}, electrodeposition\cite{dep}, and dendritic
growth\cite{dend}, among other physical processes.

Recent progress in this field has been primarily
restricted to $d=2$ dimensions, due to the availability of powerful
conformal techniques in the plane\cite{cmm}.
However, it remains important to elucidate the basic physics of DLA,
branch competition and screening, for all dimensions $d$.  A phenomenological
model of branch competition was put forth by Halsey and coworkers\cite{halsey}.
In this study, we present a high-dimensional
model in which many of the branched growth ideas of
Halsey may become exact.  We find that the branched growth
recursion equations are controlled by extremal fluctuations.
Finally, we comment on whether the extremal fluctuations
are important in DLA as well as in our model.

For a fractal cluster of dimension $D$, the number of particles
$N(k)$ on a length scale $k$ obeys
\be
\label{nk}
N(k) \propto k^D.
\ee
It is known that for DLA, $D\geq d-1$\cite{bbnd}; it is believed that
$D \rightarrow d-1$ as $d \rightarrow \infty$.  In order to study this
high-dimensional limit,
we choose to study diffusive and ballistic growth
on a Cayley tree.  Unlike models of diffusive growth
on Cayley trees in which the cluster grows from the root towards the
leaves\cite{early}, our clusters grow from the leaves towards
the root on a Cayley tree with $k$ generations of $2^k-1$ sites. 
The ballistic limit of this model has been considered previously\cite{bradley}.
There, the scaling behavior of $N(k)$ was found only for a modification of the
model, for which $N(k)$ behaves differently.

We observe non-trivial competition between the growing
branches, due to both geometrical and diffusive screening effects.
We find that extremal fluctuations dominate the growth, leading to
$N(k)$ much less than the compact result of $2^k$.  $N(k)$ still behaves
exponentially in $k$, so that $D$ is infinite.  By examining the behavior of 
$N(k)$ in different regimes, we find power law corrections to the
exponential behavior, which may be interpreted as codimensions.
\section{The Model}
Consider a Cayley tree with a constant branching factor of 2
and a specified number of levels, $k$.
We choose to initially occupy all the leaves of the tree, leaving all the 
other nodes empty.  
Therefore, a tree of $k$ levels will have a total of $2^k-1$ nodes, of
which $2^{k-1}-1$ are initially empty.
Random walkers will be released sequentially from the root of the 
tree to perform a random walk on the tree.  When a random walker moves 
onto an occupied site, it is added to the cluster at the site which it
occupied previous to the last random walk step.

Since each node has
two daughters and one parent, there is an entropic bias to the random walk,
causing it to drift towards the leaves.  We offset this bias by defining a
parameter $\gamma$ such that the walker moves towards a given daughter
node with probability $1/(2+\gamma)$, and moves towards the parent node
with probability $\gamma/(2+\gamma)$.  
At $\gamma=2$, the walker is twice as likely to move to the parent as 
to any given daughter node, so the total probability of moving
towards the leaves is equal to the probability of moving towards the root.
At $\gamma=0$, the walker only moves towards the daughter nodes, so
the growth is ballistic, not diffusive.
Growing branches can screen each other geometrically for all $\gamma$, but 
diffusive screening is present only for $\gamma>0$ (see figure 1).

Let us define the ``absorbency" of a node.
This is the probability that a random walker starting
on that node will be absorbed into the cluster without first moving
to that node's parent.  
The advantage of the absorbency is that it hides some of the non-local
effects in the calculation of random walk probabilities, and is useful
for both analytical and numerical calculations.
By definition the absorbency of an occupied node is equal to 1.

The absorbency, $a$, of a node can be obtained solely from 
the absorbencies $a_{l},a_{r}$ of its left and right
daughter nodes.  We have
\be
\label{absrec}
a=\frac{a_l+a_r}{\gamma+a_l+a_r}.
\ee
For $\gamma<2$,
we find that for a tree in the initial configuration, the absorbency of the
root tends to a constant for large $k$.  For $\gamma=2$ the initial
absorbency of the root is equal to $1/k$.
For $\gamma>2$, the absorbency of the root scales as
$\label{gg2a} (2/\gamma)^k$.  This behavior of the absorbency
suggests that $\gamma=2$ is a critical point; we confirm this
suspicion below.

Given the absorbencies, we can simulate the random
walk as follows: start with the random walker at the root.  
Move the random walker towards the leaves at each step, until it hits
an occupied node, with the 
relative probability of moving to the left or right daughters
given by the relative absorbencies. 
\section{Behavior at $\gamma=0$}
We first consider the $\gamma=0$ model, in which
the random walker only moves towards the leaves, with equal probability
of moving left and right at each stage.
Due to the branching nature of the Cayley tree, we can construct
recursion relations to describe the growth.  (These equations will be
very similar to those obtained phenomenologically in the branched
growth model.)

Define $p_0(N,k)$ to be the probability that the root is unoccupied
after $N$ particles are added to a tree of $k$ levels.  Define $p_1(N,k)$
to be the probability that the root of the tree is occupied by the
addition of the $N$-th particle.  
Define $p_2(N,k)$ to be the probability that the
root was occupied before the $N$-th particle was added (note that
this means that some of the last particles must overflow from the
tree).  Clearly,
$p_0(N,k)+p_1(N,k)+p_2(N,k)=1$.

These probabilities obey simple recursion relations.  Let $N$ particles
be added to a tree of $j+1$ levels, and assume that the root of the tree is 
unoccupied at this point.  All the particles lie below the root in
the two subtrees of the tree, and with probability
$2^{-N}{N \choose N'}$
we find that $N'$ particles were added to the left subtree and
$N-N'$ were added to the right subtree.  The root of
a given subtree
is either empty, or became occupied only as a result of the last
particle to be added to that subtree.  Therefore,
\begin{eqnarray}
\label{r1}
p_0(N,j+1) =\sum\limits_{N'=0}^{N}
2^{-N}
{N \choose N'}
\times \nonumber \\
\bigl(1-p_2(N',j)\bigr)
\bigl(1-p_2(N-N',j)\bigr).
\end{eqnarray}

Similarly, we can derive an equation for $p_1$.  We find
\begin{eqnarray}
\label{r2}
p_1(N,j+1)=\sum\limits_{N'=0}^{N}
2^{-N}
{N \choose N'}
\times \nonumber \\
2 \bigl(p_1(N'-1,j)\bigr)
\bigl(1-p_2(N-N',j)\bigr).
\end{eqnarray}
Combining the equations for $p_0$ and $p_1$, we obtain an equation
for $p_2$.

Let us make a naive guess as to the solution of these equations.  The
guess, while incorrect, will be instructive.
The number of particles which go into each subtree will fluctuate.
If we imagine that for $N$ large
these fluctuations become negligible, and further assume 
that $p_1<<p_0$, we find
that
\be
\label{noiseless}
p_0(N,j+1)=
p_0(N/2,j)^2.
\ee
The solution to this equation is $p_0(N,j)=f(N/2^j)$ such
that $f(x)=1$ for $x<x_c$ and $f(x)=0$ for $x>x_c$.  This means that
the root of the tree is always empty if less than $x_c 2^k$ particles are added
and after this the root is always occupied.

Indeed, for $j$ large, the noise in the recursion relation is exponentially
small, so eq. (\ref{noiseless}) is not so different from eq.
(\ref{r1}).  However, if we perturb $f(x)$
to $f(x)=1-\epsilon(x)$ for $x<x_c$, we find that
$\epsilon$ grows exponentially under eq. (\ref{noiseless}).  From this
we can guess that extremal statistics will become important.  Since each
node has an exponentially large number of nodes below it on the tree,
even exponentially rare events may become important.
In fact, using an argument based on extremal statistics,
we now show that the average number of particles that must be added to
a tree of $k$ levels in order to occupy the root is given by
$N(k) \approx 2^k 2^{-\sqrt{2 k}} \sqrt{k}$.

Consider a tree of $k$ levels.  Let $k=l+m$, and assume $l>>1$, $m>>1$.  
We can imagine the
tree as being made up of $2^l$ subtrees, each of $m$ levels.  Let us add
$(m-1) 2^l$ particles to the tree.  At this point, each of the subtrees 
has received roughly $m-1$ particles (later, we will consider the effects of
fluctuations in this number of particles).  
It is impossible for a subtree to
complete its growth with only $m-2$ particles added, since the subtree is $m$
levels deep, with the leaves initially seeded.
However, it is possible
for one of these subtrees to complete its growth
after only $m-1$ particles are added.
To compute this probability we note
that the first particle added to the subtree can go anywhere; the second
particle must be added to the parent node of the first particle, which
occurs with probability $2^{-(m-2)}$; the third particle must go to the
parent node of the second particle, with probability $2^{-(m-3)}$; and
so on.  This yields the result that the filling probability, which is
$p_1(m-1,m)$, is $2^{-(m-1)(m-2)/2}\approx 2^{-m^2/2}$.

Thus, if $2^l 2^{-m^2/2}$ is of order unity, it is likely that one of the
subtrees will have been filled after only $(m-1) 2^l$ particles are added to
the main tree.  At this point, the next particle added from the root has
probability $2^{-l}$ of attaching to the parent of the filled subtree, so that
it will take approximately $2^l$ further particles added
before the parent node of the
subtree is occupied.  At that point, it will take approximately $2^{l-1}$ 
more particles before the grandparent of the subtree is occupied.  
Eventually, after $2^l+2^{l-1}+2^{l-2}+...=2^{l+1}$ particles are added,
we occupy the root of the whole tree.   At this point, the total number
of particles added is $(m-1)2^l +2^{l+1}\approx m 2^l$.

If we pick $l$ and $m$ correctly,
we will thus take only $m 2^l$ particles to fill the tree.  Since 
$2^l 2^{-m^2/2}$
must be of order unity, $l \approx m^2/2$.  
For $k>>1$, we find $m \approx\sqrt{2 k}$ and $l\approx k-\sqrt{2 k}$.
Once this number ($m 2^l$) particles has been added, we expect the tree
to be filled so that
\be
\label{bnd}
N(k) \lesssim m 2^l\approx 2^k 2^{-\sqrt{2 k}} \sqrt{2k}.
\ee

Had we chosen $l,m$ such that $l>m^2/2$
we would find many filled subtrees, and a weaker inequality than eq.
(\ref{bnd}).
For $l<m^2/2$, the argument
does not work, because no subtree is filled.

Now, suppose we add $\tilde N (k)<<2^k 2^{-\sqrt{2 k}} \sqrt{2k}$ particles.  
If we pick $l$ and $m$
as above with $m\approx\sqrt{2k}$, and if each of the subtrees receives of order
$\tilde N(k)/2^l<<m$ particles, then it is unlikely that any of the $2^l$
subtrees gets completed, hence the whole tree can not be completed.
Thus if each subtree receives roughly $N(k)/2^l$ particles, eq.
(\ref{bnd}) is an estimate and not just an inequality.

Neglecting the effects of geometrical screening
between subtrees, the number of particles 
arriving in a subtree follows Poissonian statistics.  Thus, while
each subtree will receive roughly $N(k)/2^l$ particles, there
are rare events in which a subtree receives a large number of particles.  
Including these fluctuations, one can show that the probability that a 
given subtree becomes occupied is
\be
\label{poisson}
\sum\limits_{n=1}^{\infty} p_1(n,m) \frac{n_0^n}{n!}e^{-n_0},
\ee
with $n_0=N(k)/2^l$.  The number of particles arriving in each
subtree is still of order $N(k)/2^l$ and the inclusion of these fluctuations
does not change the estimate of the order of magnitude of $N(k)$;
the rare events considered in which a typical number of particles
arrive in a subtree and form an atypical structure are much more likely than
events in which an exponentially larger number of particles arrive and
form a typical structure.

To justify the neglect of geometrical screening, we note that
for $l,m$, such that $l<m^2/2$, we expect at most one
subtree is occupied.  Thus, the argument is self-consistent.

Finally, let us express these estimates in terms of the recursion 
relations for $p_0,p_1$.  First, we have found that 
$p_0(N,m)=1$ for $N<m-1$.  However,
\be
\label{extremal}
p_0(m-1,m) = 1-{\rm O}(2^{-m^2/2}).
\ee
Next, we have used eq. (\ref{noiseless})
to argue that $p_0(2^l(m-1),m+l) = 1-{\rm O}(2^l 2^{-m^2/2})$
so that for $2^l 2^{-m^2/2}\approx 1$, the probability $p_0(2^l(m-1),m+l)<<1$,
which indicates that the root is likely to be occupied.
Using eq. (\ref{r1}) instead of eq. (\ref{noiseless}) corresponds
to including the fluctuations expressed by eq. (\ref{poisson}).
\section{Behavior at $\gamma>0$}
At $\gamma=0$, 
the only screening effect is the geometrical screening effect.
We now repeat
the argument based on extremal statistics for the case $\gamma>0$, when
diffusive effects begin to increase the competition between subtrees.  

For $\gamma>0$ it is impossible to write closed recursion
relations for the probabilities $p_0,p_1,p_2$. 
We can, however, write
more general recursion relations, which include the absorbencies of the nodes.

Still, for any $\gamma$, it is impossible to reach the root of an $m$ level
tree with $m-2$ particles.  
The probability of reaching the root of an $m$ level tree with
$m-1$ particles can be computed as in the $\gamma=0$ case, now
taking the absorbencies into account.  For $\gamma<2$, the
result is still given by eq. (\ref{extremal}), while for
$\gamma=2$, we find
$
p_0((m-1),m) = 1-{\rm O}(m! \,2^{-m^2/2}),
$
and for $\gamma>2$, 
$
p_0((m-1),m) = 1-{\rm O}(2^{-m^2/(2 {\rm log}_2\gamma}))$.

Using the same extremal argument as above, we find that, for $\gamma<2$,
\be
\label{gl2}
N(k) \lesssim 2^k 2^{-\sqrt {2 k}} \sqrt{2k},
\ee
for $\gamma=2$,
\be
\label{ge2}
N(k)\lesssim 2^k 2^{-\sqrt {2 k}},
\ee
and for $\gamma>2$,
\be
\label{gg2}
N(k)\lesssim 2^k 2^{-\sqrt {2 k {\rm log}_2{\gamma} }} 
\sqrt{2 k {\rm log}_2
\gamma}.
\ee
In deriving this, we assume that each of the $2^l$ subtrees continues to
receive roughly $N/2^l$ particles.  However, now the number of particles 
arriving
in a given subtree becomes non-Poissonian due to the effects of diffusive
screening.  Physically, this effect must reduce $N(k)$.

We still conjecture that eqs. (\ref{gl2},\ref{ge2},\ref{gg2}) are
estimates, and not just inequalities.
Suppose we add only $\tilde N(k)<<m 2^l$ particles to the main tree, with $m,l$ 
chosen to obtain the most stringent inequality.  
Let us show that this is not enough particles to grow
to the root of the tree.  We again do this by showing that it is 
justified to assume that each subtree receives roughly $\tilde N(k)/2^l$ 
particles, 
so that none of the subtrees can be completely occupied.  To justify
this assumption,
we now need to show that diffusive, as well as geometrical,
screening effects are negligible.  When a
subtree has received roughly $\tilde N(k)/2^l<<m$ particles, the
top $O(m)$ levels of each subtree are still likely to be completely empty.
As a result, the absorbency of the subtree is 
is only slightly changed from that of a completely
empty subtree and so the
assumption of Poissonian statistics is reasonable.

Interestingly, the $\gamma=2$ result for $N(k)$ is reduced from the
$\gamma<2$ result by an amount of order $\sqrt{k}$, implying
a codimension of $1/2$ at the critical point.
For $\gamma>2$, we find an infinite codimension, as the corrections to the 
$\gamma<2$ results are exponential, not power law.
These results should be compared to the expected co-dimension of
1 for DLA in high dimensions.
\section{Comparison to Numerics}
Using the absorbency technique, we have performed numerical calculations on 
the model for trees up to $k=24$ for $\gamma=0,1,2,4$.
In figure 2 we show ${\rm log_2}(N(k)/2^k)$ averaged
over 100 runs for each data point.  As can be
seen, $N(k)<<2^k$.  Next, in figure 3, 
we show $N(k)$ divided by the appropriate theoretical result for the 
scaling of $N(k)$.  For all curves, we see that the plot decreases
initially, and then appears to stabilize at some given value, indicating
that the theoretical result is in agreement with numerics.
The theoretical calculation is only able to produce the asymptotic scaling
of $N(k)$.  However, the prefactor for $N(k)$
appears to be close to unity for $\gamma=0$.
\section{Conclusion}
We have introduced a model that
provides an interesting realization of some ideas of branched
growth.  We find in eq. (\ref{noiseless}) a simple recursion
equation for the branch competition.  However, the solution of this
equation is unstable to noise.  
On the Cayley tree, the noise grows indefinitely
leading to a situation controlled by extremal statistics, not unlike
the ``infinite disorder" critical points known in statistical 
mechanics\cite{dma}.  

In finite dimensional DLA, we still expect
that branch competition will ``self-generate" noise, causing small
perturbations of a noiseless solution to grow, leading to a regime with
non-vanishing fluctuations.  Extremal statistics have been used to
estimate the fractal dimension of DLA by Turkevich and Scher\cite{turk}.  
The relative importance of large fluctuations 
as a function of dimensionality remains an unsolved problem.

\noindent *current address: CNLS, MS B258, Los Alamos National
Laboratory, Los Alamos, NM 87545, hastings@cnls.lanl.gov

\begin{figure}[!t]
\begin{center}
\leavevmode
\epsfig{figure=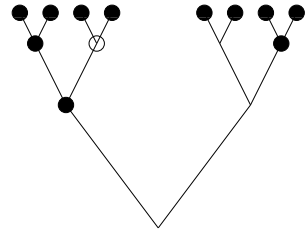,height=3cm,angle=0}
\end{center}
\caption{
Screening for a tree with $k=4$ after three growth steps.
Filled circles represent occupied nodes.
The empty circle represents a node that
is geometrically screened with no growth possible on this site.  For
$\gamma>0$, there is a diffusive screening effect:
the next random walker starting from the root
is more likely to occupy the root as a
result of a move onto the occupied left daughter node than it is
to be added to the right half of the tree.
}
\label{fig1}
\end{figure}
\begin{figure}[!t]
\begin{center}
\leavevmode
\epsfig{figure=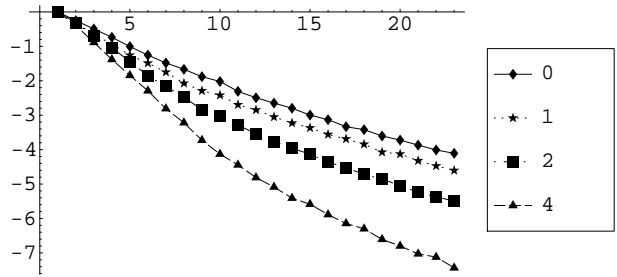,height=4cm,angle=0}
\end{center}
\caption{${\rm log_2} (N(k)/2^k)$ as a function of $k-1$ for $\gamma=0,1,2,4$}
\label{fig2}
\end{figure}
\newpage
\begin{figure}[!t]
\begin{center}
\leavevmode
\epsfig{figure=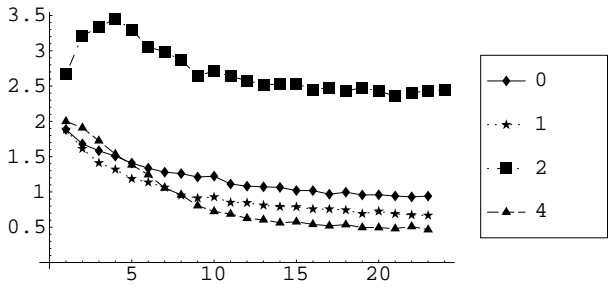,height=4cm,angle=0}
\end{center}
\caption{$N(k)/(2^k 2^{-\sqrt{2 k}}\sqrt{2 k})$
as a function of $k-1$ for $\gamma=0,1$;
$N(k)/(2^k 2^{-\sqrt{2 k}})$ for $\gamma=2$;
$N(k)/(2^k 2^{-\sqrt{2 k {\rm log}_2 \gamma}} \sqrt{2 k {\rm log}_2
\gamma})$ for $\gamma=4$.  All curves asymptote to a constant.}
\label{fig3}
\end{figure}
\end{document}